\documentclass[12pt]{iopart}

\usepackage{amssymb}
\usepackage{cite}

\usepackage{bm}
\usepackage{upgreek}
\usepackage{graphicx}

\newcommand{\be}{\begin{equation}}
\newcommand{\ee}{\end{equation}}

\begin{document}

\title[Non-perturbative dispersion interactions]{ Non-perturbative theory of dispersion interactions}

\author{M Bostr{\"o}m$^{1,2}$, P Thiyam$ ^{3}$, C
Persson$^{4,1,3}$, D F Parsons$^5$, S Y Buhmann$^6,7$, I Brevik$^2$
and Bo E Sernelius$^{8}$}

\address{$^1$Centre for Materials Science and Nanotechnology, University of Oslo, P O Box 1048 Blindern, NO-0316 Oslo, Norway}
\address{$^2$Department of Energy and Process Engineering, Norwegian University of Science and Technology, N-7491 Trondheim, Norway}
\address{$^3$Department of Materials Science and Engineering, Royal Institute of Technology, SE-100 44 Stockholm, Sweden}
\address{$^4$Department of Physics, University of Oslo, P O Box 1048 Blindern, NO-0316 Oslo, Norway}
\address{$^5$Department of Applied Mathematics, Australian National University, Canberra, Australia}
\address{$^6$Physikalisches Institut, Albert-Ludwigs-Universit{\"a}t
Freiburg, Hermann-Herder-Str.~3, 79104 Freiburg, Germany}
\address{$^6$Freiburg Institute for Advanced Studies,
Albert-Ludwigs-Universit{\"a}t Freiburg,
Albertstr.~19, 79104 Freiburg, Germany}
\address{$^8$Division of Theory and Modeling, Department of Physics,
Chemistry and Biology, Link{\"o}ping University, SE-581 83
Link{\"o}ping, Sweden}

\ead{iver.h.brevik@ntnu.no}

\pacs{
	31.30.jh, 	
	34.20.Cf, 	
	42.50.Lc, 	
	03.70.+k 	
	}

\begin{abstract}

Some open questions exist with  fluctuation-induced forces between extended dipoles. Conventional intuition derives from  large-separation perturbative approximations to dispersion force theory.  Here we present  a full non-perturbative  theory. In addition  we discuss how one can take into account finite dipole size corrections.  It is of fundamental value to investigate the limits of validity of the perturbative dispersion force theory.

\end{abstract}

\today

\maketitle

For  many years there has been strong  interest, both theoretically and experimentally, in fluctuation induced forces\,\cite{Milt,Ninhb,Pars,Buhmann,Mitchell}.
Until now most work on dispersion forces using either the quantum electrodynamical or the semi-classical formalism have relied  upon a point dipolar
description. The relationship between these  two theories has been discussed by Ninham, Parsegan, and Weiss\,\cite{NPW}. Previous work that indeed did incorporate the finite
size of an extended dipole was based on perturbative
expansions (truncated to the leading term),  suggesting that the van
der Waals energy contributed with a short-range attractive binding
energy\,\cite{Mahanty}.
Our non-perturbative approach has been  investigated for atom-atom dispersion forces in water\,\cite{Priya1} and for extended dipole self-energies.\,\cite{Priya2}
The effects of a non-perturbative theory on self-energies (which can give  50$\%$ changes or more compared to perturbative theory) ought in principle to be detectable experimentally, although it is a challenge to measure them directly. Solvation energies of atoms in a dielectric medium, i.e. changes of the self-energy in a vacuum compared to in a medium, can be measured\,\cite{Honig,Tim}.   Use of an incorrect theory may lead to problems in interpretations of experimental solvation energies. Conceptually important additional effects of the non-perturbative theory will be demonstrated in the present work for dipole-dipole dispersion interactions in vacuum.
Perturbative expansions
in terms of multiple dipole-field interactions are often restricted
to leading-order approximations where each dipole scatters the field
only once.
This is a valid approach if the coupling is very weak and/or the
separation between the dipoles is large enough.
The aim of this work is to demonstrate how the keeping of the full
non-perturbative, multi-scattering
theory and taking the finite spread of the dipole into account,  alters
the non-retarded van der Waals force at contact distances. The inclusion of  finite dipole size  in a
non-perturbative theory opens up for the possibility of having van der
Waals repulsion when two dipoles come very close to each other.

In the following we first recapitulate the theory of van der Waals energy between point like dipoles.
Then we  introduce a theory that accounts for finite spread of dipoles in the van der Waals force between two dipoles in their ground states. The van der Waals potential is created by a sum of symmetric and antisymmetric modes. These modes can be related to the non-retarded first order dispersion forces.\,\cite{Bostrom1} Repulsion in the  van der Waals force may occur as some modes stop contributing when the dipoles come close together. Within this theory avoiding perturbative series expansions, we find substantial corrections to the dipole self-energy.\,\cite{Priya2} We also consider the resonance interaction when one dipole is in its ground state and the other is in an excited state. At resonance the excitation switches back and forth between the two dipoles.   The interaction can then be separated into three branches, the $j=x$-, $y$-, and $z$-branches, where the name of each branch denotes in what direction the oscillating dipoles are pointing. We let the $z$-direction be defined as the line joining the two dipoles.
We focus on deriving analytical asymptotes that can be used for all separations within the non-retarded regime.
The correct way to evaluate resonance interaction  in the retarded limit is still debated\,\cite{Bostrom1,McLachlan}. We present results in the non-retarded limit. It is reassuring that   in this limit different groups obtain the same results.\,\cite{Bostrom1,McLachlan} As we will demonstrate when two dipoles come close together the results remain finite in contrast to the textbook results (which is a $1/\rho^3$-dependence for resonance interaction and $1/\rho^6$-dependence for ground state van der Waals potentials where $\rho$ is the interdipole separation).

Mitchell et al. published in 1972 a paper \cite{Mitchell} demonstrating short-range repulsion between two point-like dipoles when they come very close together. Consider two identical dipoles with polarizability given by

\begin{equation}
\alpha (i \omega)=\alpha_0/(1+\omega^2/\omega_0^2),
\label{Eq1a}
\end{equation}
with $\omega_0$ being the ionization frequency of the dipole.  The allowed coupled frequencies in the limit of small separations are given by Eq. (27) in Mitchell et al.\,\cite{Mitchell}:
\begin{equation}
\omega=\omega_0 \sqrt{1 \pm \alpha_0/\rho^3}, \omega=\omega_0 \sqrt{1 \pm 2 \alpha_0/\rho^3},
\label{Eq2a}
\end{equation}
$\rho$ being the distance between the dipoles.
The first two roots each occur twice. An important point is  that  only real frequencies contribute to the interaction energy. If all roots are real, i.e. $2 \alpha_0/\rho^3<1$, then the interaction energy is

\begin{equation}
\begin{array}{l}
U(\rho ) = \frac{{\hbar {\omega _0}}}{2}\\
\quad  \times \left\{ {2\sqrt {1 + {\alpha _0}/{\rho ^3}}  + 2\sqrt {1 - {\alpha _0}/{\rho ^3}} } \right.\\
\left. {\quad  + \sqrt {1 + 2{\alpha _0}/{\rho ^3}}  + \sqrt {1 - 2{\alpha _0}/{\rho ^3}}  - 6} \right\}.
\end{array}
\label{Eq3a}
\end{equation}

For large distances this expression goes over to the London interaction energy.  The  key point here as demonstrated by Mitchell et al.\,\cite{Mitchell} was that, for decreasing separations, first $2 \alpha_0/\rho^3>1$ and then $\alpha_0/\rho^3>1$, several modes become imaginary and do not contribute to the (real) energy. The interaction energy becomes in this limit
\begin{equation}
\begin{array}{l}
U(\rho ) = \frac{{\hbar {\omega _0}}}{2}\left\{ {2\sqrt {1 + {\alpha _0}/{\rho ^3}}  + \sqrt {1 + 2{\alpha _0}/{\rho ^3}}  - 6} \right\}\\
\quad \quad \quad  \approx \frac{{\hbar {\omega _0}}}{2}\left\{ {(2 + \sqrt 2 )\sqrt {{\alpha _0}} /{\rho ^{3/2}} - 6} \right\}.
\end{array}
\label{Eq4a}
\end{equation}

The freezing out of modes at distances of the order of the dimensions of the dipoles  leads to a repulsive energy.  Of course the dipole approximation itself breaks down in this limit.

This problem is partially relaxed by including spread out size effects of the extended dipole. A difficulty still occurs, however,  as can be seen from the fact that the logarithm in the usual expression for the interaction energy between two identical molecules, for example,    turns imaginary when $x>1$ for some of the Matsubara frequencies. One has to   sum over terms  of the form $\ln(1-x)$. The problem will be addressed in the following discussion.

From the equations of motion for the excited system it is straightforward to derive the zero temperature Green's function for two identical dipoles in vacuum\,\cite{McLachlan,Bostrom1}. The resonance condition\,\cite{Bostrom1} can be obtained from the following relation: $\tilde 1 - \alpha (\omega )^2 {\tilde T}(\rho |\omega )^2 = 0$, where $\tilde 1$ is the identity tensor,  $\rho$ is the dipole-dipole separation distance, $\tilde T$ is the susceptibility tensor and   $\alpha (\omega)$  the polarizability (at frequency $\omega$) of the dipole.
In the case of two identical dipoles excited in the $j$th branch the above resonance condition can be separated into one antisymmetric and one symmetric part. Here we refer to the quantum mechanical state function. In the symmetric case the state function is the sum of the function  where dipole 1 is excited and dipole 2 not and the function where dipole 2 is excited and dipole 1 not. In the antisymmetric case the state function is the difference between these two states. Since the excited symmetric state has a much shorter life time than the antisymmetric state the system can be trapped in an excited antisymmetric state\,\cite{McLachlan}.   The resonance interaction energy of this antisymmetric state can be evaluated by a simple  expression for  two identical dipoles,\,\cite{Bostrom1}
\begin{equation}
{U_j}(\rho ) = \hbar \int\limits_{ - \infty }^\infty  {\frac{{d\xi }}{{2\pi }}} \ln \left[ {1 + \alpha (i\xi ){T_{jj}}(\rho |i\xi )} \right].
\label{Eq1}
\end{equation}

The corresponding van der Waals (Casimir-Polder) interaction between identical dipoles in their ground states is given by the following expression:
\begin{equation}
{U_{CP}}(\rho ) = \frac{\hbar }{2}\sum\limits_{j = x,y,z} {\int\limits_{ - \infty }^\infty  {\frac{{d\xi }}{{2\pi }}} \ln \left[ {1 - \alpha {{(i\xi )}^2}{T_{jj}}{{(\rho |i\xi )}^2}} \right]}
\label{Eq2}
\end{equation}

Let us pause here to discuss the interaction energy in terms of electromagnetic normal modes. The interaction energy at a specific separation is the sum of the zero point energy of all modes minus the corresponding sum at infinite separation. Some of the modes are attractive and some repulsive. Most of the interaction is canceled out but a net attraction survives. When the dipoles are brought closer together the energies of the attractive modes decrease while those of the repulsive modes increase. In the formalism associated to each mode corresponds one with minus the same energy; these negative-energy modes do not contribute to the interaction. When the dipoles come close enough together the energy of the most attractive mode approaches  zero value.
When this happens the one with minus the same energy reaches zero from the negative side. With further reduction of the separation the two modes move away, one along the positive and one along the negative imaginary axis. This may happen for several modes before the dipoles are in direct contact with each other. Our interpretation is that when this happens the modes stop contributing to the interaction. Since it is only attractive modes that drop out, the attraction gets weaker and the net interaction can become repulsive.

With our interpretation one could perform the calculation in one of three equivalent ways: One is the mode-summation method. This method can only be used in rare cases when the distinct modes can be found. Then the energy of the modes that reach the zero value should be put equal to zero.
This holds when the separation continues to decrease. Another way is to proceed as we did above. Then the integration should be performed not at the imaginary axis but along a vertical  line an infinitesimal distance from and to the right of the axis. The result is real valued since the imaginary part of the integrand is  odd and the real part an even function of $\xi$. A third way is to integrate along a closed path encircling the whole positive real axis. All three ways give the same result.

\bigskip
\noindent{\it Perturbative theory.}

The traditional way to treat the integrands of Eqs.\,(\ref{Eq1}) and (\ref{Eq2}) is to make a series expansion of the logarithm and keep the lowest order term only, i.e. $\ln (1 + x) \approx x$. This is permissible for large separations when the lowest order term is small compared to unity. In this section, we consider this perturbative case while in the next section, we refrain from doing this approximation and term it the non-perturbative case. The second improvement we do is to take the finite size of the dipoles into account instead of treating the dipoles as point particles.
The polarization cloud of real dipoles has a finite spread and we consider as an interesting case a spatial distribution of the dipole following a Gaussian function\,\cite{Mahanty3}.  We have, following the formalism developed by Mahanty and Ninham\,\cite{Mahanty3}, derived the Green's function elements that account for retardation, background media, and finite dipole Gaussian radius ($a$) in an accurate way.
We now consider the effect of the finite dipolar size on the non-retarded resonance interaction between two dipoles in an excited state in vacuum.  In this limit (velocity of light $\to \infty $) we can use the non-retarded Green's function elements,
\begin{equation}
\begin{array}{*{20}{l}}
{T_{xx}^{NR}(\rho |i\xi ) = T_{yy}^{NR}(\rho |i\xi )}\\
\quad\quad\quad\quad\quad{ = \frac{{ - 1}}{{\sqrt \pi  { \rho ^3}}}\left[ {\sqrt \pi  {\rm{erf}}\left( {\frac{\rho }{a}} \right) - 2\left( {\frac{\rho }{a}} \right){e^{ - {{\left( {\frac{\rho }{a}} \right)}^2}}}} \right],}
\end{array}
\label{Eq3}
\end{equation}
and
\begin{equation}
\begin{array}{*{20}{l}}
{T_{zz}^{NR}(\rho |i\xi )}\\
\quad\quad{ = \frac{{ - 2}}{{\sqrt \pi  {\rho ^3}}}\left[ { - \sqrt \pi  {\rm{erf}}\left( {\frac{\rho }{a}} \right) + 2\left( {\frac{\rho }{a}} \right)\left[ {1 + {{\left( {\frac{\rho }{a}} \right)}^2}} \right]{e^{ - {{\left( {\frac{\rho }{a}} \right)}^2}}}} \right].}
\end{array}
\label{Eq4}
\end{equation}

We begin by studying the effect of the finite size when we expand the
logarithm and truncate to leading order in the dipole-field interaction.
We find for the
resonance interaction with the $x$-branch excited the following long
range result in the non-retarded limit,
\begin{equation}
\begin{array}{*{20}{l}}
{U_x^{NR}(\rho) \cong  - [2\hbar /(\sqrt \pi  {\rho ^3})]}\\
{\quad \quad \quad  \times \left[ {\sqrt \pi  {\rm{erf}}\left( {\frac{\rho }{a}} \right) - 2\left( {\frac{\rho }{a}} \right){e^{ - {{\left( {\frac{\rho }{a}} \right)}^2}}}} \right]\int\limits_0^\infty  {\frac{{d\xi }}{{2\pi }}} \alpha (i\xi ).}
\end{array}
\label{Eq5}
\end{equation}
The corresponding result  with the $z$ branch excited is,
\begin{equation}
\begin{array}{*{20}{l}}
{U_z^{NR}(\rho) \cong \left[ {4\hbar /(\sqrt \pi  {\rho ^3})} \right]\left\{ {\sqrt \pi  {\rm{erf}}\left( {\frac{\rho }{a}} \right)} \right.}\\
{\quad \quad \quad \left. { - 2\left( {\frac{\rho }{a}} \right)\left[ {1 + {{\left( {\frac{\rho }{a}} \right)}^2}} \right]{e^{ - {{\left( {\frac{\rho }{a}} \right)}^2}}}} \right\}\int\limits_0^\infty  {\frac{{d\xi }}{{2\pi }}} \alpha (i\xi ).}
\end{array}
\label{Eq6}
\end{equation}

It is of interest to present the corresponding results at close contact.  Taking $t=\rho/a$ the $\rho$ dependence for the $x$-branch is contained in
$f({t}) =  [ \sqrt{\pi} \rm{erf( t)} - 2 t \exp(-t^2) ]/{t}^3$.
The small $t$ expansion of the error function is ${\rm{erf}}(t) = (2/\sqrt \pi  )\exp ( - {t^2})(t + 2{t^3}/3 + O[{t^5}])$,
and therefore $f(t) = 4/3 + O[t^2]$. This produces the limit $T_{xx} \rightarrow -4 / (  3 \sqrt{\pi}  a^3  )$. One can in the same way show that
$T_{zz} \rightarrow T_{xx} $. Therefore in the limit of small dipole-dipole separation we find
$U_z^{NR}=U_x^{NR} \to \frac{{-8\hbar }}{{3\sqrt \pi  {a^3}}}\int\limits_0^\infty  {\frac{{d\xi }}{{2\pi }}} \alpha (i\xi )$.
As a result, unlike the well known long distance behavior
\cite{McLachlan}, when two dipoles come very close together the
resonance interaction is the same for all excitation-branches. We
observe that the resonance interaction at close contact depends on the
radius $\propto a^{-3}$. This should be compared to
the corresponding result for ground state van der Waals interaction
that depends on the radius $\propto a^{-6}$. The analytical asymptotes
are only valid for dipolar systems
where the perturbative, single-scattering expansion exists.
Finite size effects
soften the attraction at very small separations.  For isotropically
excited dipole pairs the first expansion term in the logarithm cancels
out and the leading non-retarded term is $\propto a^{-6}$.

\bigskip
\noindent{\it Non-perturbative theory}

Now we move to the results of the non-perturbative theory. The complete expressions for the non-retarded first order dispersion potentials follow trivially when Eq.\,(\ref{Eq1})  is combined with Eq.\,(\ref{Eq3}) and Eq.\,(\ref{Eq4}).
The simple expressions found for the finite size non-retarded resonance interaction suggest that one can solve this problem analytically, at least if one chooses a simple model  polarizability ($\alpha (i\xi)=\alpha(0)/(1+\xi^2/\omega_0^2)$). We then find the following expression (valid when the dipoles are close enough together so that retardation effects can be neglected):
\begin{equation}
\begin{array}{*{20}{l}}
{U_j^{NR}(\rho ) = \frac{{\hbar {\omega _0}}}{\pi }\int_0^\infty  d x\ln \left[ {\left| {1 + \frac{{\alpha (0)T_{jj}^{NR}}}{{1 + {x^2}}}} \right|} \right]}\\
{ = \hbar {\omega _0}{\mathop{\rm Re}\nolimits} \left[ { - 1 + \sqrt {1 + \alpha (0)T_{jj}^{NR}} } \right].}
\end{array}
\label{Eq7}
\end{equation}

The non-retarded Casimir-Polder potential between two dipoles with finite radii is

\begin{equation}
\begin{array}{*{20}{l}}
{U_{CP}^{NR}(\rho ) = \frac{{\hbar {\omega _0}}}{2}\sum\limits_{{{j = x}},{{y}},{{z}}} {{\mathop{\rm Re}\nolimits} \left[ { - 2 + \sqrt {1 + \alpha (0)T_{jj}^{NR}} } \right.} }\\
{\left. { + \sqrt {1 - \alpha (0)T_{jj}^{NR}} } \right].}
\end{array}
\label{Eq8}
\end{equation}
Asymptotically  the interaction approaches the long range non-retarded limit with a $\rho^{-6}$-dependence.
In the limit of zero separation between  two dipoles we obtain,
\begin{equation}
\begin{array}{l}
U_{CP}^{NR}(\rho  = 0) = \frac{{3\hbar {\omega _0}}}{2}{\rm{Re}}\left[ { - 2 + \sqrt {1 + 4\alpha (0)/(3\sqrt \pi  {a^3})} } \right.\\
\left. { + \sqrt {1 - 4\alpha (0)/(3\sqrt \pi  {a^3})} } \right].
\end{array}
\label{Eq8b}
\end{equation}
This energy limit for two overlapping polarization clouds is made up of a bonding and an antibonding self-energy contribution which should be compared with the self-energy of a single dipole given later in this work. While the van der Waals repulsion between point like dipoles goes towards infinity the previous expression provides the upper limit to the van der Waals repulsion between finite sized dipoles.   This dipole-dipole contribution is part of the ``attractive'' term in the Lennard-Jones potential. Our simple expression for the non-retarded van der Waals potential can easily be included in simulations that are using Lennard-Jones potentials. Only real modes contribute to the interaction energy\,\cite{Mitchell}.
It is important to note that besides the attractive term one should also add the repulsive term in the Lennard-Jones potential corresponding to
exchange interactions due to wave-function overlap and  higher-order multipole transitions\,\cite{r0a, r0b, r0c}. For the perturbative case, using the formalism of Richardson \,\cite{Richardson} we found in an earlier work \,\cite{Priya1} that the dipole-dipole interaction gives larger contribution than the dipole-quadrupole and quadrupole-quadrupole interactions at all separations. Multipolar
contributions may, however,  provide as much as 50$\%$ of the dispersion energy of some atoms, electron-rich anions in particular\,\cite{Tim}. They have been
found necessary for reproducing experimental properties such as ion
solvation energies \,\cite{r1}, entropies and partial molar volumes \,\cite{r2},
surface tensions \,\cite{r3} and activity coefficients \,\cite{r4}.
 Finite size corrections prevent the attraction from going to infinity when two
dipoles come very close together.  In the limit of small dipole-dipole separations the
van der Waals interaction from the full expressions may turn into a
constant repulsive energy when the polarization clouds overlap, but there are also effects for two dipoles at contact distances that reduce the attractive binding energy.
The interaction approaches asymptotically the well-known
attractive  $\rho^{-6}$-dependence.
Perturbative expansions have also been used to calculate the dispersion
self-energy of a single dipole in vaccum.\,\cite{Mahanty}. A more
complete theory gives the following dispersion self-energy\,\cite{Priya2},
\begin{equation}
\begin{array}{l}
{U_S} = \frac{{3\hbar {\omega _0}}}{2}\left[ { - 1 + \sqrt {1 + 4\alpha (0)/(3{a^3}\sqrt \pi  )} } \right].
\end{array}
\label{Eq9}
\end{equation}
\begin{table}
\caption{The finite size van der Waals binding energy,
$U_{CP}^{NR}(\rho=0)$ using the expression derived here, and as comparison the dispersion self-energy\,\cite{Priya2},
$U_{\rm{S}}$, for noble gas atoms. The superscript 'truncated' indicates
that the result is from using a truncated (or perturbative)
expansion of the logarithm in the integrand.  All energies are
in eV.}
\label{values}
\begin{center}
\begin{tabular}{cccccc}
\hline
\hline
\multicolumn{1}{c}{Element} &\multicolumn{1}{c}{$U_{{\rm{CP}}}^{{\rm{full}}}\left( 0 \right)$} &\multicolumn{1}{c}{$U_{{\rm{CP}}}^{{\rm{truncated}}}\left( 0 \right)$} &\multicolumn{1}{c}{$U_{\rm{S}}^{{\rm{full}}}$} &\multicolumn{1}{c}{$U_{\rm{S}}^{{\rm{truncated}}}$}\\
\hline
He&29.06 &-409.6 &71.21 &131.53\\
Ne&58.20 &-1046 &104.95 &220.28\\
Ar&8.767 &-133.7 &37.56 &62.07\\
Kr&4.366 &-87.98 &29.94 &47.50\\
\hline
\hline
\end{tabular}
\end{center}
\end{table}

When series expanding and truncating this expression we can rederive the dispersion self-energy found by Mahanty\,\cite{Mahanty}. However, the validity of a series expansion assumes that $\alpha(0)/a^3$ is much smaller than unity which is not always the case. We provide in Table I the finite size van der Waals binding energy and the finite size dispersion self-energy for a number of atoms.   The input data were taken from Refs.\,\cite{atompaper,Mahan,Priya2}.
We give the results found when using our non-perturbative theory and as comparison also the approximate theory using a  truncated series.  The corrections to the self-energy of atoms are substantial and measurable. In fact, the non-perturbative self-energies in our examples are up to a factor of 2 different from the results from the truncated theory.
These effects of a non-perturbative theory ought in principle to be detectable experimentally.  They should be accounted for to avoid incorrect interpretations of experimental solvation energies and permeabilities across membranes.

\bigskip
\noindent{\it Conclusion}

We conclude that finite size effects keep the interaction finite when two dipoles come close together. Notably the resonance interactions between two identical dipoles are the same, and attractive, for $x$-, $y$-, and $z$-excitation branches when the two atoms come close enough together. The  Casimir-Polder interaction energy which is attractive at long range approaches a constant repulsive van der Waals energy at short range. At distances beyond a few dipole radii the different interactions approach well known results from perturbation theory.  The non-perturbative theory is conceptually important, but it is also important for the binding energy of dipole pairs and self-energies. We emphasize that the reason why the interaction stays finite when the dipoles come close to each other is the spead out of the polarization cloud.

There are a number of potential applications for a non-perturbative theory of intermolecular forces.
Take for example the solvation energies of ions in a dielectric medium (i.e. changes of the self-energy in a vacuum compared to in a medium) that can be measured\,\cite{Honig,Tim}. Latimer et al\,\cite{Latimer} were able to fit experimental solvation energies to the Born equation by increasing the effective radius of the ions.  Self-energy changes have also been shown to influence permeabilities across membranes \,\cite{Pars1969}.   The results obtained for  the permeability of atoms across a membrane will be changed  due to changes in the self-energy in vacuum.  A factor of 2 difference for these self-energies compared to those  obtained from a  series-expanded theory ought  therefore  to be measurable with existing experimental equipment for solvation free energies and permeabilities.

 \bigskip

 MB and CP acknowledge support from the Research Council of Norway (Project: 221469).  PT acknowledges support from the European Commission.
SYB gratefully acknowledges support by the German Research Council
(grant BU 1803/3-1) and the Freiburg Institute for Advanced Studies.
CP acknowledges support from Swedish Research Council (Contract No.
C0485101).

\end{document}